\documentclass[twocolumn,groupedaddress,
superscriptaddress,preprintnumbers,
amsmath,amssymb,
aps,final
]{revtex4-2}
\usepackage{graphicx}
\usepackage{dcolumn}
\usepackage{bm}%
\usepackage[colorlinks=true,citecolor=blue]{hyperref}
\DeclareUnicodeCharacter{2032}{\ensuremath{'}}
\hypersetup{colorlinks=true,citecolor=blue,linkcolor=red,urlcolor=blue}

\usepackage{float}
\usepackage{amsmath}

\usepackage{bm}
\newcommand{\beq}{\begin{equation}}
\newcommand{\eeq}{\end{equation}}
\newcommand{\beqnar}{\begin{eqnarray}}
\newcommand{\eeqnar}{\end{eqnarray}}
\newcommand{\bfig}{\begin{figure}}
\newcommand{\efig}{\end{figure}}

\usepackage{color}
\usepackage[dvipsnames]{xcolor} 
\usepackage{changes} 

\begin{document}

\title{Modulating dichroism and optical conductivity in bilayer graphene under intense electromagnetic field irradiation}
\author{S. Sajad Dabiri}
\affiliation{Department of Physics, Shahid Beheshti University
, 1983969411 Tehran, Iran}
\author{Hosein Cheraghchi}
\affiliation{School of Physics, Damghan University, 36716-41167, Damghan, Iran}
\affiliation{School of Physics, Institute for Research i n Fundamental Sciences (IPM), Tehran 19395-5531, Iran}
\author{Fatemeh Adinehvand}
\affiliation{School of Physics, Damghan University, 36716-41167, Damghan, Iran}
\author{Reza Asgari}
\affiliation{Department of Physics, Zhejiang Normal University, Jinhua 321004, China}
\affiliation{School of Physics, Institute for Research in Fundamental Sciences (IPM), Tehran 19395-5531, Iran}

\begin{abstract}
This study explores the impact of a strong perpendicular laser field on the electronic structure and optical conductivity of bilayer graphene. Employing the Floquet-Bloch theorem and a four-band Hamiltonian model, we calculate the optical conductivity, unveiling modified optical properties due to the altered band structure.
We investigate the effects of both circularly and linearly polarized dressing fields on the electronic structure and optical conductivity in the system. Under linear polarization, we observe a notable anisotropy in the band dispersion and optical conductivity, resulting in linear dichroism. In the case of circular polarization, we anticipate the emergence of induced Berry curvature and circular dichroism, especially close to the dynamical gaps. When circularly polarized light is applied alongside a bias potential, the band structure differs for right-handed and left-handed polarization. In this case, the longitudinal optical conductivity remains the same for both, while the transversal optical conductivity exhibits distinct results.
Furthermore, the induced Berry curvature and valley asymmetry introduce the potential for generating a valley-polarized current, enabling valley-selective pumping and leading to circular dichroism.
\end{abstract}

\maketitle
\section{Introduction}
The interaction between time-periodic light and materials is a fundamental and pivotal area of research in modern optics and photonics, holding broad implications for a diverse range of technological applications. Over the past decade, significant focus has been dedicated to engineering electronic bands through the interaction of Bloch states with time-periodic fields~\cite{oka2019}. This interaction allows for precise control over the features of the band structure by modulating the polarization, intensity, and frequency of the monochromatic dressing field. The Floquet-Bloch (FB) theory extends Bloch's theorem, originally concerned with spatial periodicity, to incorporate time periodicity. This means that when spatial periodicities in lattices interact with time-periodic fields, the Hilbert space expands to encompass FB states~\cite{shirley65,hanggi88}. This extension is central to comprehending the behavior of electrons in crystalline solids subjected to time-varying fields. Furthermore, it establishes a versatile framework for manipulating the spectrum through photo-induced resonances.

Recent advancements in technology have brought forth intense high-frequency (mid-infrared range) laser light as a dressing field, allowing the study of photo-dressed electrons influenced by polarized electromagnetic fields. The modification of band structures due to electron-photon interactions has been explored in various quantum systems, including topological insulators \cite{2013Wang}, ultra-cold atom gases \cite{2}, graphene structures \cite{3}, HgTe/CdTe quantum wells \cite{2013Cayssol,2011Lindner}, \added{and topological insulator thin films \cite{dabiri1,dabiri2}}, to name a few. Furthermore, the application of mid-infrared laser fields on materials has enabled the realization of novel phenomena, such as Floquet topological insulators \cite{2011Lindner,2010Kitagawa}. Intriguingly, this illumination not only leads to the appearance of topologically protected edge states along irradiated graphene sheets but also around large defects and adatoms within the dynamical gap \cite{2016Lovey}.

Experimentally, FB states have been studied in irradiated topological insulator Bi$_2$Se$_3$, where the gapless surface states follow a Dirac cone spectrum \cite{2013Wang,2016Mahmood}. Time-and-angle resolved photo-emission spectroscopy (Tr-ARPES) proves to be a potent technique for resolving such photo-induced band gaps, making the measurement of FB states feasible. These states manifest as replicas of the original Dirac cone in the surface states of irradiated topological insulators \cite{2013Wang}, and their dynamical gap (the photo-induced gap) depends on the polarization and intensity of the laser field \cite{2013Wang}.

Predictions are abound for various phenomena in irradiated graphene, including the photovoltaic Hall effect \cite{2009Oka}, a metal-insulator transition \cite{2010Kibis}, transport through n-p junctions under electromagnetic fields \cite{2008Syzranov,2007Fistul}, and a photo-induced quantum Hall effect even in the absence of a magnetic field \cite{2011kitagawa}. \added{Furthermore, simulation of Floquet topological insulators in classical systems was realized for example in photonic systems \cite{photonicfti}, acoustic systems \cite{acousticfti}, and proposed in electrical circuits \cite{dabiri2023}.}

 Extensive efforts have been devoted to manipulating and creating a band gap in the band structure of monolayer graphene, which inherently exhibits a gapless semimetallic behavior. Notably, it has been demonstrated that linearly polarized light induces an anisotropic gapless band dispersion in graphene, while circularly polarized fields open an isotropic gap \cite{2016Kristinsson}. Additionally, a dynamical gap emerges in graphene at $ \Omega/2$, where $\Omega$ represents the frequency of the laser field when irradiated by circularly polarized mid-infrared laser fields \cite{2009Oka,2008Syzranov,2014Usaj}. Proposals have also been made regarding the merging and shifting of Dirac points in driven graphene \cite{2013Scholz,2013Delplace}. Twisted bilayer graphene, characterized by its large unit cell, manifests extraordinary electronic flat bands at magic twist angles, giving rise to correlated electronic states. The combination of exceptionally flat bands with a non-zero Chern number, along with gaps and correlated states, positions these materials as an intriguing platform for investigating the interplay between topology and interaction. The introduction of circularly polarized light to such materials allows for the manipulation of the topological properties of these flat bands \cite{seradjeh, sentef}, a control that is influenced by both the twist angle and the perpendicular applied bias.

Optical conductivity, on the other hand, is a material property that describes how well the material conducts electric current in response to an applied electromagnetic field, particularly in the optical frequency range. The optical conductivity of materials interacting with time-periodic light fields provides a fundamental understanding of how these materials respond to electromagnetic radiation. This knowledge is essential for a wide range of applications in modern technology and materials science. Therefore, optical conductivity plays a crucial role in experiments and, as such warrants necessitate careful calculation \cite{2009Oka,2008Syzranov}. In irradiated graphene, a multi-step-like structure in optical conductivity has been reported, originating from optical transitions involving FB states  \cite{2011Zhou,2013Scholz}. These FB states are contingent on the polarization, intensity, and frequency of the laser field.

Bilayer graphene (BLG) stands out as a highly versatile and promising material due to its combination of a tunable band gap, valley physics, and topological effects. This makes it an excellent platform for exploring novel phenomena in the presence of laser fields. By applying time-periodic external fields (Floquet engineering), BLG's electronic properties can be finely tuned, potentially leading to the emergence of new electronic states, topological phases, and exotic phenomena. With its unique characteristics, including a tunable band gap induced by a perpendicular applied bias, valley Hall effect, and quadratic band touching points, BLG offers a rich arena for investigating the influence of laser irradiation. For instance, the application of an off-resonant linearly polarized light (LPL) on BLG can induce a Lifshitz transition near the Fermi level \cite{2017Iorsh}. Additionally, studies have revealed that irradiation with left and right-handed circularly polarized electromagnetic fields can break valley symmetry in gated BLG \cite{2010Abergel,2017Kibis}. However, several intriguing questions persist. What is the impact of irradiation with different polarizations on the time-averaged optical conductivity of BLG? How does the occupation of Floquet states impact optical conductivity? Additionally, could we elaborate on the significance of the valley trace in optical conductivity as a measurable physical quantity? Does dichroism arise for different dressing field frequencies? These queries to name a few continue to be captivating subjects for this study. We illustrate that the application of bias disrupts the valley symmetry in the dressing spectrum induced by either left- or right-handed circularly polarized light. When BLG is simultaneously subjected to circularly polarized light and a perpendicular bias voltage, there is no equivalence observed in the spectrum surrounding the K and K' valleys. Right-handed and left-handed circularly polarized light exhibit distinct quasi-energy spectra, while the longitudinal optical conductivity remains consistent for both polarizations. However, disparities arise in the transverse optical conductivity. In contrast, linear polarized light induces linear circularly polarized light.

The paper is organized as follows.  The formalism provides the Floquet-Kubo formula and the optical conductivity of driven BLG in Sec. \ref{Sec1}. We initially delve into the influence of off-resonant high-frequency light with different polarizations on the band structure and phase diagram of BLG, considering the additional influence of bias potential in Sec. \ref{Sec2}. Following this, we examine the effect of on-resonant light on the system's band structure and optical conductivity. Given the significance of the occupation of Floquet states in this regime, we compare the quench occupation of Floquet states with the mean-energy occupation. Additionally, we demonstrate how valley asymmetry in the presence of bias potential leads to intriguing effects in both frequency regimes of the dressing field. Finally, we wrap our main results in Sec. \ref{Sec4}.

\section{Theory and Model}\label{Sec1}

We will commence our investigation with the study of BLG in the presence of a bias potential. Subsequently, \added{we will present the Floquet-Kubo formula for time-averaged optical conductivity obtained from the Floquet theory}. Within this framework, we will analyze the key parameters like quasi-energy, the phase diagram stemming from band gap closing in which the valley-polarized Hall insulator with the zero Chern number and a quantum anomalous Hall insulator with a Chern number of $\pm4$ emerge, and then we calculate the optical conductivity in both off- and on-resonant regimes.

Bernal-stacked BLG comprises two honeycomb layers: an upper ($u$) layer and a lower ($d$) layer of carbon atoms. These layers are rotated by $\pi/6$ concerning each other, causing $A_{u}$ and $B_{d}$ atoms to align directly atop one another. In the tight-binding model, we designate $\gamma_{0}$ as the hopping energy among intra-layer nearest neighbor sites, and $\gamma_1$ as the hopping energy among inter-layer nearest neighbor sites ($A_{u}-B_{d}$). Therefore, the Hamiltonian \cite{2006McCann}, particularly near the $\xi=\pm1$ valley, can be expressed as follows: 
\begin{equation}
\hat{H}={\xi} \begin{pmatrix}
V/2 & 0 & 0& v_{F}{\bf \pi}^{\dagger}\\
0& -V/2 & v_{F}{\bf \pi} & 0 \\
0& v_{F}{\bf \pi}^{\dagger}&-V/2 & {\xi} \gamma_{1}\\
v_{F}{\bf \pi} & 0 &{\xi} \gamma_{1}& V/2
\end{pmatrix}
\label{H0}
\end{equation}

We choose $\psi=(A_{d}, B_{u} ,A_{u}, B_{d})^{T} $ for $\xi=+1$ and $\psi=(B_{u}, A_{d} ,B_{d}, A_{u})^{T}$ for $\xi=-1$ as the basis sets in the Hamiltonian, and define ${\bf \pi}=(k_{x}+ik_{y})$. Here, $k_x$ and $k_y$ represent the wave vectors of electrons in the $x$ and $y$ directions, respectively. The velocity attributed to the coupling is denoted as $v_{F}=\sqrt{3}a\gamma_{0}/2$, where $a=0.246$ nm stands for the lattice constant. In addition, in Eq.~(\ref{H0}), $V$ represents the bias potential applied perpendicular to the plane of BLG. The parameters are assigned the values of $\gamma_{0}=3.16$ eV and $ \gamma_{1}=0.390$ eV~\cite{2009Kuzmenko}. The energy dispersion of the Hamiltonian in Eq.~(\ref{H0}) can be obtained as:
\begin{equation}
\begin{aligned}
E_{\eta}^2&=\dfrac{\gamma_1^2}{2}+\dfrac{V^2}{4}+v_F^2 k^2+\eta \sqrt{C},\\
C&=\frac{\gamma_1^4}{4}+v_F^2k^2[\gamma_1^2+V^2]
\end{aligned}
\label{E}
\end{equation}
where $k= \sqrt{k_x^2+k_y^2}$ and $\theta=\arctan(k_y/k_x)$. Here, $\eta=\pm1 $ denotes different energy bands.

The foundational tools for Floquet engineering have been outlined in Refs. \cite{handbook, oka2019}. Let us briefly discuss some key aspects of the Floquet theorem. By illuminating the BLG plane with an intense laser field, the time-dependent field can be incorporated into the Hamiltonian as follows (assuming $\hbar=1$): $ {\bf k}\rightarrow {\bf k}+e {\bf A}(t) $, where the vector potential takes the form of $ {\bf A}^{c}(t)=A_{0}(\sin \Omega t,\cos \Omega t )$ for circular polarization, and $ {\bf A}^{l}(t)=A_{0}( 0,\cos \Omega t )$ for linear polarization. Here, $\Omega$ represents the frequency of the laser field. For right-handed (left-handed) polarization, $\Omega$ should be taken as positive (negative). It is important to note that since the vector potential is independent of position, the wave vector $k$ remains a conserved quantum number \cite{2013Scholz}. 

\added{The essential tools derived from the Floquet theorem are presented in Appendix~\ref{ftheorem}. These include the definition of Floquet quasi modes $|\phi_\nu(t)\rangle$, quasi energies $\varepsilon_{\nu}$, Fourier components of quasi modes $|\phi^n_\nu\rangle$, and the Floquet-Schr\"{o}dinger equation in the extended Hilbert space. The physical weights and mean energy of Floquet bands are precisely defined in Eqs.~(\ref{weight}) and (\ref{menergy}), respectively. Additionally, the time-averaged density of states is introduced and defined in Eq. (\ref{DOSF}).}

The study of optical properties in solids is a well-explored area, encompassing both experimental and theoretical investigations~\cite{girvin2019modern}. The interaction with light can alter the quantum characteristics of matter, including its conductivity, which in turn facilitates the development of optically driven devices. Beyond applications in devices, the material's optical response offers a potent means to investigate the quantum states of electrons and their excitations within spatially periodic potentials~\cite{cupo2023optical}. Therefore, it is essential to calculate the optical properties of the studied system. The real component of the time-averaged optical conductivity within the linear response theory can be derived through non-equilibrium Green's function as the following \cite{2011Zhou,kumar,dabiri3}
\begin{equation}
\begin{aligned}
&\text{Re} {\bar{\sigma}_{ll}(\omega)}=\dfrac{\pi g_s}{\omega}\sum_{\textbf{k},m}\sum_{\nu<\mu} 
\vert j^{l(m)}_{\nu\mu}\vert^{2}(f_{\nu}-f_{\mu})\times\\
&\big[\delta(\omega+\epsilon_{\nu}-\epsilon_{\mu}-m\Omega)-\delta(\omega-(\epsilon_{\nu}-\epsilon_{\mu}-m\Omega))\big]
\end{aligned}
\label{oc}
\end{equation}
Here, the current matrix elements are defined as $j^{(m)}_{\nu\mu}=\dfrac{1}{T}\int_{0}^{T} \langle\phi_{\nu}(t)\vert j(t) \vert\phi_{\mu}(t) \rangle e^{im\Omega t}dt$ and $\delta$ stands for the Dirac delta function. If the current operator $j=e \partial H (t)/\partial k_l$ is independent of time, $j^{(m)}_{\nu\mu}=\sum_n \langle \phi_{\nu}^{(n)}\vert j \vert\phi_{\mu}^{(n+m)} \rangle $. Also, $f_\nu$ is the occupation of energy band $\nu$ which depends on switch-on protocols and relaxation mechanisms and would not be derived easily according to the Fermi-Dirac distribution function like non-driven systems. It should be remembered that the above summations (Eq.(\ref{oc})) would run over $\bold{k}$ around both valleys $K$ and $K'$. As we will show later, there is no equivalence in the spectrum around the $K$ and $K'$ valleys when circularly polarized light and a perpendicular bias voltage are simultaneously applied on BLG. 

Next, we introduce the expression for the imaginary component of the time-averaged optical Hall conductivity in periodically driven systems. This can be obtained as: \cite{fiete2016,2011Zhou,kumar,dabiri3} 
\begin{equation}
\begin{aligned}
&\text{Im} {\bar{\sigma}_{ll'}(\omega)}=g_s \sum_{\textbf{k},m}\sum_{\nu<\mu} 
2 \text{Im}[ j^{l(m)}_{\nu\mu} j^{l'(-m)}_{\mu\nu}]  (f_{\nu}-f_{\mu})\times\\
&\big[\frac{2\omega\eta_0}{(\omega^2-(\epsilon_{\nu}-\epsilon_{\mu}-m\Omega)^2)^2+4\omega^2\eta_0^2}\big]
\end{aligned}
\label{ochall0}
\end{equation}
where $\eta_0$ is an infinitesimal constant. At zero-temperature and tiny $\eta_0$, the imaginary component of the time-averaged optical Hall conductivity is given by 
\begin{equation}
\begin{aligned}
&\text{Im} {\bar{\sigma}_{ll'}(\omega)}=\dfrac{\pi g_s}{\omega}\sum_{\textbf{k},m}\sum_{\nu<\mu} 
 \text{Im}[ j^{l(m)}_{\nu\mu} j^{l'(-m)}_{\mu\nu}]  (f_{\nu}-f_{\mu})\times\\
&\big[\delta(\omega+\epsilon_{\nu}-\epsilon_{\mu}-m\Omega)+\delta(\omega-(\epsilon_{\nu}-\epsilon_{\mu}-m\Omega))\big]
\end{aligned}
\label{ochall}
\end{equation}
The imaginary component of the optical Hall conductivity quantifies the absorption of circularly polarized probe light. Right and left-handed circularly polarized light exhibit distinct absorption patterns, corresponding to $\text{Re}\sigma_{xx}\mp\text{Im}\sigma_{xy}$, respectively. Nonzero values of the imaginary part of the transverse optical conductivity suggest the presence of circular dichroism. Furthermore, the existence of Berry curvature in the band structure can be deduced from circular dichroism and, consequently, from $\text{Im}\sigma_{xy}(\omega)$ (see the Appendix).

Regarding Eq.~(\ref{ochall}), it is noteworthy that at a fixed frequency of the probe light $\omega$, the contribution to $\text{Im}\sigma_{xy}$ arises exclusively from optical transitions between Floquet states whose energy difference equals $\omega$ in the quasi-energy spectrum. This implies that integrating over a small momentum space range allows the calculation of optical conductivity at low probe frequencies, justifying the use of a low-energy $\mathbf{k} \cdot {\bf p}$ Hamiltonian.

We should emphasize that the components of the frequency-dependent permittivity tensor of the studied system can also be computed using the relation $\epsilon_{ll'}(\omega)=\delta_{ll'}+i \frac{1}{\epsilon_0\omega} \sigma_{ll'}(\omega)$, provided that the optical conductivity has been calculated. Here, $\epsilon_0$ is known as vacuum permittivity.

\section{Numerical Results and Discussions}\label{Sec2}
In the subsequent subsections, we first examine the case of off-resonant driving, where the frequency of the driving light exceeds the bandwidth. Following that, we delve into the impact of on-resonant illumination with a low frequency.

In our numerical calculations, we employ a Gaussian function, $\delta(x) = \frac{1}{\eta \sqrt{2\pi}}e^{-x^2/2\eta^2}$, as an approximation for the delta function appearing in the DOS and optical conductivity and we hence set $\eta=1$. Furthermore, all conductivity curves are normalized by $\sigma_0={e^2}/{2}$.


As demonstrated in Eq.~(\ref{weight}), the contribution of each Floquet-Bloch state in the band structure is determined by the weight coefficient ($W_k^n$ for the $n^{th}$ band) associated with the Floquet-Bloch wave function ($\phi^{n}_{\nu}$). Taking this into consideration, the integration range for calculating the conductivity is chosen as -0.55 nm$^{-1} < k < 0.55 $nm$^{-1}$, with $n_k\approx1000$ representing the number of $k$-steps. The allowed optical transitions and weight coefficient effectively limit the contribution of large $k$-values to physical quantities such as DOS and conductivity, particularly in the low-energy range.
Throughout this section, we consider a Floquet Hamiltonian dimension of $n_F=(2n_t+1)n_{\text{cell}}$, where $n_t=15$ and $n_{\text{cell}}=4$.

\subsection{Off-resonant Regime}\label{Sec2-1}
In the case of off-resonant driving, the physical weights of the bands in the first Floquet zone ($- \Omega /2<\varepsilon_\nu<\Omega/2$) are much higher than the other sidebands, particularly if the amplitude of the drive is sufficiently low. Reference~\cite{2011kitagawa} demonstrates that for the off-resonant drive, $|\phi^n\rangle\approx\frac{H^{(n)}}{n\Omega}|\phi^0\rangle$ for $n\neq0$ where $H^{(n)}$ is defined in Eq.~(\ref{ht}) of Appendix. We assume the dimensionless parameter $\frac{H^{(n)}}{n\Omega}\ll1$ for $n\neq1$ to be in the weak driving regime, where the central sideband has the most significant effect.

Subsequently, one can concentrate on the zeroth sideband and consider the impact of other sidebands perturbatively. It is possible to employ an effective Hamiltonian that describes the system's evolution at integer multiples of the time period. Expansions in inverse powers of frequency for the effective Hamiltonian exist, which can be truncated for sufficiently low amplitudes of drive~\cite{Bukov,eckart,bw}.
\begin{equation}
H^{\text{eff}}=H^{(0)}+\sum_{n}(n \Omega)^{-1}[H^{(-n)},H^{(+n)}]+O(1/(\Omega)^2)
\label{photon_H}
\end{equation}
\subsubsection{Off-resonant: Circular polarization}

The phase diagram in the Hartree-Fock approximation, along with the presence of other effects like valley exchange interaction, was presented in \cite{2dmat}.
Now, let us briefly touch upon the irradiation effects on the system described by Hamiltonian Eq.~(\ref{H0}). After applying Peierls' substitution to Eq.~(\ref{H0}) and computing the Fourier components, one can determine the effective Hamiltonian for circular polarization, which is  
\begin{equation}
\begin{aligned}
H^{\text{eff}}=H^{(0)}+ \Delta_\Omega v_F^2 \mathbf{1}\otimes\sigma_z
\label{Heff}
\end{aligned}
\end{equation}
where $H^{(0)}$ is equal to Eq.~(\ref{H0}) and $\mathbf{1},\sigma_z$ are identity and Pauli matrices, respectively, and $\Delta_\Omega=\mathcal{A}^2/\Omega$ with $\mathcal{A}=e A_0$.

The band gap given by Eq.~(\ref{Heff}) is equal to
\begin{equation}
\frac{\text{gap}}{2}=\Delta_\Omega v_F^2+\frac{V}{2} \xi
\label{gap}
\end{equation}

\begin{figure}
\includegraphics[width=\linewidth]{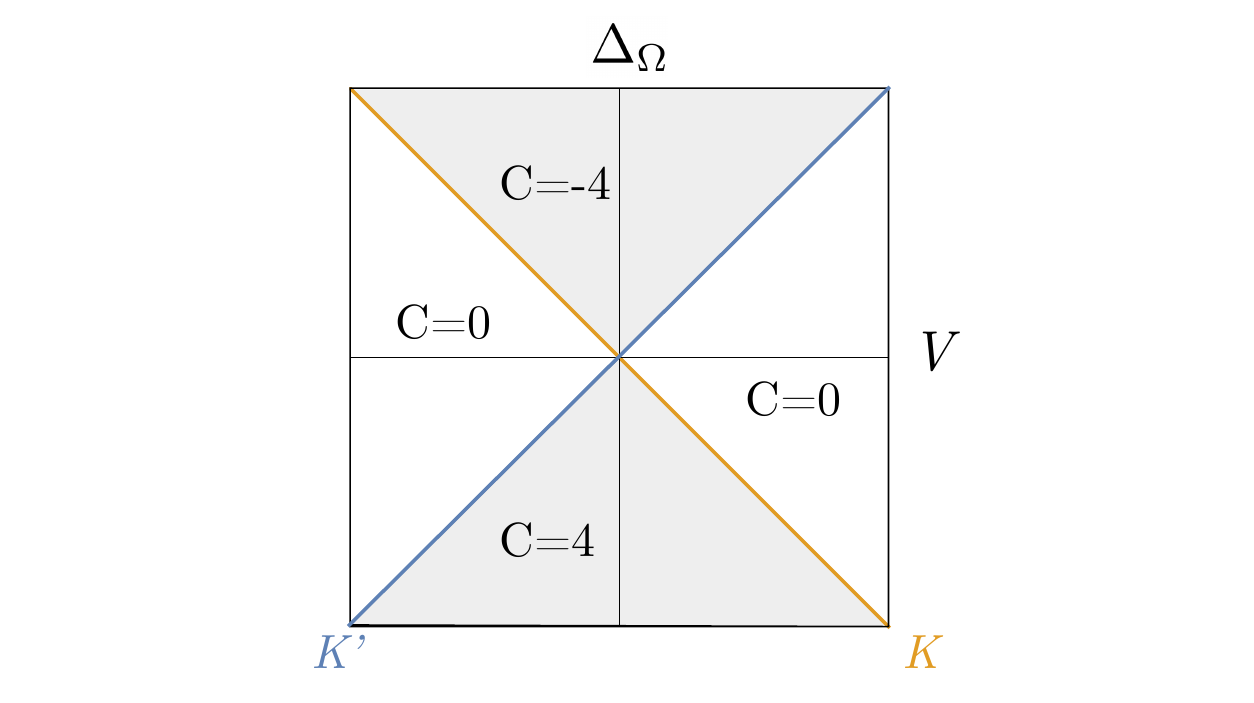}
\caption{(Color online) The phase diagram in the presence of off-resonant drive in the studied system. The grey and white regions show the quantum anomalous Hall insulator and quantum valley polarized Hall insulator, respectively. The orange and blue lines show the gap closing at K and $K^{'}$ valley respectively. Notice that the asymmetry between K and $K^{'}$ opens up the possibility of generating a valley-polarized current by tuning the chemical potential and applying a voltage to the system.}
\label{Fig1}
\end{figure}

We confirm that the gap closing can only occur at the $\Gamma$ point, which can be deduced from Eq.~(\ref{gap}). The resulting phase diagram is illustrated in Fig.~\ref{Fig1}. Notice that the Chern number depicted in the phase diagram of Fig.~\ref{Fig1} is computed using the numerical method introduced by Fukui, et al. \cite{FHS}. This method is applied to the effective Hamiltonian described in Eq.~(\ref{Heff}). 
The white regions represent the valley-polarized Hall insulator, where the Chern number is zero, but edge modes from different valleys propagate helically. The grey regions correspond to a quantum anomalous Hall insulator with a Chern number of $\pm4$. 
The gap in the spectrum at the $K$ valley differs from that at the $K'$ valley, which is expected from Eq.~(\ref{gap}). In Fig.~\ref{Fig1}, the gap closing at the $K$ and $K'$ valleys is indicated by orange and blue colors, respectively. This asymmetry between the two valleys opens up the possibility of generating a valley-polarized current by tuning the chemical potential and also applying a bias voltage to the system. Additionally, a probe light can be directed towards the sample and selectively absorbed by one valley, a phenomenon known as \emph{valley-selective pumping}.
On the other hand, we anticipate a \emph{circular dichroism} (i.e., different absorbance for right and left-handed polarized probe light) due to the Berry curvature induced by light \cite{2dmat,opto}. In the presence of Berry curvature, when the probe frequency is set to the value of the gap at each valley, nearly complete circular dichroism is expected, meaning one of the right-handed or left-handed polarized light is not absorbed \cite{2dmat,opto}. However, in the absence of $\Delta_\Omega$ and the presence of $V$, there is no complete circular dichroism because both valleys have the same magnitude of gap and opposite Berry curvatures.
It is worth noting that a similar phase diagram emerges when we introduce magnetization instead of light \cite{qahbilayer}. In this case, however, there is no asymmetry between valleys.

\begin{figure}
\includegraphics[width=\linewidth]{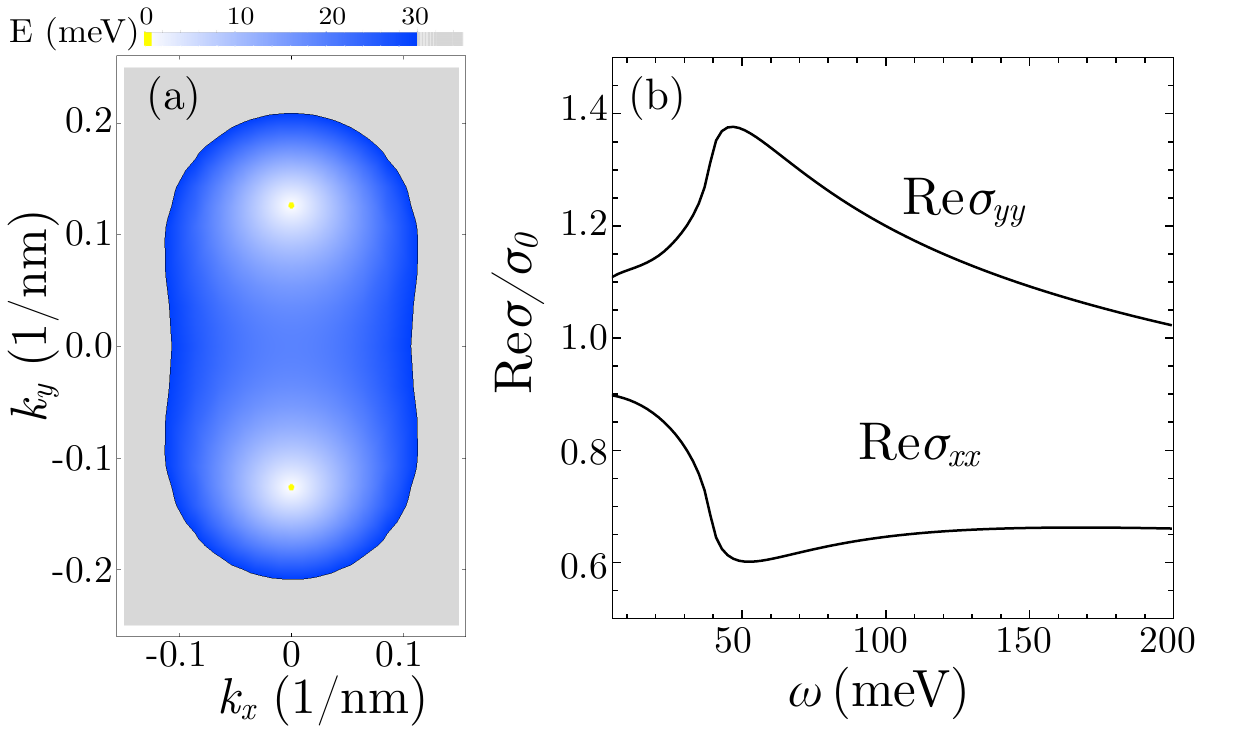}
\caption{(Color online) (a) Quasi-energies of irradiated BLG by LPL showing two gapless points. (b) The real part of the time-averaged longitudinal optical conductivity for the system. Notice that $Re \sigma_{xx}\neq Re\sigma_{yy}$ due to band gap asymmetry around K and $K^{'}$ and hence it turns out that there is a linear dichroism in the presence of linearly polarized dressing field. We set frequency being $\Omega=1$~eV and intensity $\mathcal{A'}=0.1/v_F^2$ along the $y$ direction.}
\label{Fig2}
\end{figure}

\subsubsection{Off-resonant: Linear polarization}

For linearly polarized light (LPL), the second term in Eq.~(\ref{photon_H}) vanishes, and it becomes necessary to calculate the second-order contribution. Since for linear polarization, $H^{(1)}=H^{(-1)}$ and $H^{(n)}=0$ for $n\neq 0,\pm1$, we can apply the van Vleck expansion \cite{bw} and thus we have
\begin{equation}
\begin{aligned}
H^{\text{eff}}_{\text{lin}}&=H^{(0)}+\frac{[H^{(1)},[H^{(0)},H^{(1)}]]}{\Omega^2}
\label{HlvV}
\end{aligned}
\end{equation}
Assuming the polarization is along the $y$ axis, we can write the effective Hamiltonian as
\begin{equation}
\begin{aligned}
H^{\text{eff}}_{\text{lin}}&=H^{(0)}+\frac{\mathcal{A}^2}{\Omega^2}\left(-\xi  k_x v_F^3 \sigma_x\otimes\sigma_x -\frac{1}{2}v_F^2\gamma_1 \mathbf{1}\otimes\sigma_x \right)
\label{Hlin}
\end{aligned}
\end{equation}

Let us investigate how the band structure of Eq.~(\ref{Hlin}) changes as a function of its parameters. First, in the absence of $V$, the system is always gapless. 
We assume $\mathcal{A'}=\mathcal{A}^2 /\Omega^2<v_F^{-2}$ because Eq.~(\ref{Hlin}) is not applicable for large amplitudes of drive.
 So, for the aforementioned condition, the gap closing can occur just at $(0,k^*_y)$ where ${k^*_y}^2=- \mathcal{A'} (-2+\mathcal{A'} v_F^2)\gamma_1^2/{4  }$.

We show the energy dispersion and longitudinal optical conductivity of BLG irradiated by LPL along the $y$ axis in Fig.~\ref{Fig2}. The frequency of dressing field light is $\Omega=1$~eV and the intensity $\mathcal{A'}=0.1v_F^{-2}$. This figure is obtained by using the effective Hamiltonian in Eq.~(\ref{Hlin}) and static Kubo Formula which has good agreement with the results by using Floquet Hamiltonian and dynamical Kubo Formula Eq.~(\ref{oc}). Two gapless points are visible in Fig.~\ref{Fig2} (a) and Fig.~\ref{Fig2} (b) implying that there is a \emph{linear dichroism} (i.e. a difference between conductivity along two perpendicular directions) in the presence of linearly polarized dressing field.

It should be mentioned that by adding the bias potential $V$ and for $\mathcal{A'} <v_F^{-2}$ the system is always gapped.

We examine the impact of the trigonal warping (a correction which leads to four gapless points on the Fermi level \cite{2009Kuzmenko}) in the system and we conclude that the above argument is qualitatively applicable. For instance, in the absence of $V$, by applying an LPL in every direction, the system remains gapless but just two gapless points along the direction of polarization only exist for high enough amplitudes. Actually, in the absence of the dressing field, there are four gapless points in the system, but turning on the dressing field makes them merge and by increasing the intensity of the dressing field, finally, two of them remain. This point was also noted in Ref.~\cite{2017Kibis} where the direction of polarization is assumed to be along the $y-$axis. It was shown that after the application of light, two of the four energy pockets near the Fermi level merge and become gapped, but the Chern number remains constant \cite{2017Kibis}. For other directions of polarization of the laser, two gapless points remain for high enough amplitudes.

\subsection{On-resonant regime}\label{Sec2-2}
In this subsection, we present the numerical results for the Floquet band structure and optical conductivity under on-resonant driving conditions. 

\begin{figure}
\includegraphics[width=\linewidth]{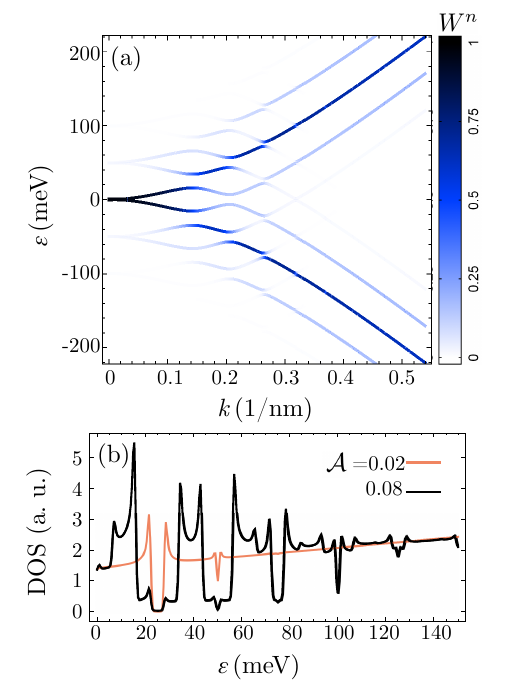}
\caption{(a) Quasi-energies of irradiated BLG with CPL with frequency of $\Omega=50$~meV and intensity $\mathcal{A}=0.08$~1/nm. (b) Time-averaged DOS for two intensities $\mathcal{A}=0.02, 0.08$~1/nm. \added{Color scale in part (a) shows the physical weights $W^n_\nu$ defined in Eq.~(\ref{weight}).}}
\label{Fig3}
\end{figure}

\subsubsection{On-resonant: Circular polarization}

In Fig.~\ref{Fig3}(a), we present the Floquet band structure of irradiated BLG subjected to circularly polarized light (CPL) with a frequency of $\Omega=50$meV and intensity $\mathcal{A}=0.08$~(1/nm). The color scale represents the physical weights as defined in Eq.~(\ref{weight}). While there are four Floquet bands in each Floquet zone, it is evident that two of them carry significant weights within the quasienergy range depicted in the figure. This implies that a two-band low-energy Hamiltonian can be used instead of the full four-band Hamiltonian when focusing on low-energy properties. However, in this paper, we utilize the complete four-band Hamiltonian. The band structure reveals dynamical gaps at quasienergies $n\Omega/2$, and as expected, for larger values of $\mathbf{k}$, the gap size decreases.

In Fig.~\ref{Fig3}(b), we present the time-averaged density of states (DOS) of the system, calculated using Eq.~(\ref{DOSF}). The light frequency is the same as in panel (a), and two intensities, namely $\mathcal{A}=0.02$ and $\mathcal{A}=0.08$~(1/nm), are plotted. This figure also showcases the development of gaps in the DOS, which corresponds to the gaps seen in the band structure. Furthermore, it can be inferred that higher intensities lead to more van Hove singularities in the DOS. This is because as the field intensity increases, more bands acquire significant weights.


To calculate the time-averaged optical conductivity, we first need to determine the occupation of Floquet states $f_\nu$ as defined in Eq.~(\ref{oc}). In this paper, we employ the quench occupation model, often referred to as the "sudden approximation" \cite{dehghani,2011Zhou}. This model assumes that the system was initially in the ground state of the non-driven model, and the drive is suddenly turned on. The occupation of states is obtained by projecting the non-driven ground state onto the Floquet wave function, yielding $f_\nu=\sum_{\mu,n}n_F(E_\mu)|\langle g_\mu|\phi_\nu^{(n)}\rangle|^2$. This differs from the mean-energy assumption \cite{2011Zhou,dabiri3}, where each Floquet state is occupied according to its mean energy (at zero temperature, $ f^{mean}_\nu=\Theta(E_F-\bar{\epsilon}_\nu $).

\begin{figure}
\includegraphics[width=\linewidth]{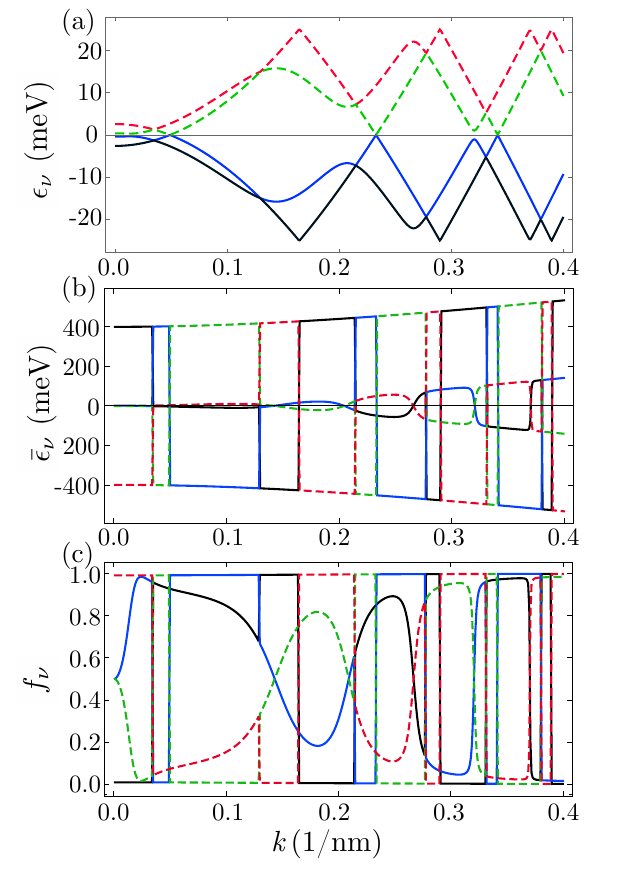}
\caption{ (Color online) (a) quasienergy bands in the first Floquet-Brillouin zone (b) Mean-energy and (c) quench occupation of Floquet states as a function of the momentum for irradiated BLG with CPL. Each colored curve indicates one Floquet band ($\nu=1,2,3,4$) of the four bands in the first Floquet zone.  We set frequency to $\Omega=50$~meV and intensity to $\mathcal{A}=0.08$ nm$^{-1}$. }
\label{Fig4}
\end{figure}

It turns out that the presence of time-periodic light can lead to the creation of additional energy levels, known as sidebands, in the material's electronic structure. These sidebands result from the interaction between electrons and photons and play a crucial role in the material's response to the applied field. To explore the relationship between the two occupation models, we present in Fig.~\ref{Fig4} the quasienergy bands in the first Floquet-Brillouin zone (i.e. (-$\Omega/2,\Omega/2$)) (a), mean energy (b) and quench occupation of states (c). It is evident from these figures that at each resonance between Floquet bands, the mean energy and occupation of states interchange. Notably, one band can resonate with two other bands. Another noteworthy feature is that when the mean energy is positive (negative), the quench occupation is lower (higher) than 0.5. Since the difference between the occupation of states (the factor $f_\nu-f_\mu$ in Eq.~(\ref{oc}) for low values of $k$ is smaller than the mean-energy occupation case (which is always 0 or $\pm1$), it is expected that the magnitude of optical conductivity for lower frequencies with quench occupation is, in most cases, lower than the mean-energy occupation (see Fig. 8 of Ref. \cite{2011Zhou}). However, exceptions may exist in some frequency regions. Nevertheless, the fundamental characteristics of the two models remain consistent.



\begin{figure}
\centering
\includegraphics[width=\linewidth]{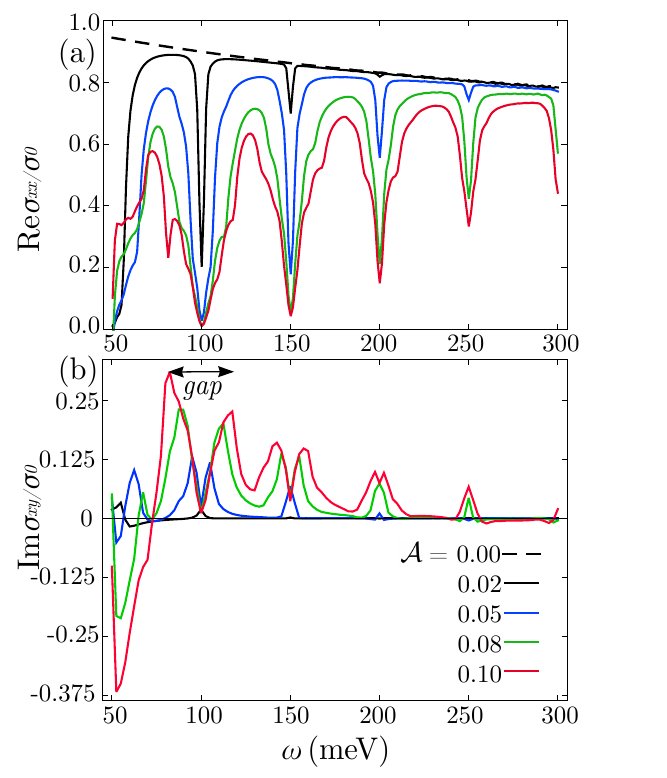}
\caption{(Color online) a) Real part of time-averaged longitudinal optical conductivity, b) Imaginary part of optical Hall conductivity as a function of the probe frequency for a BLG irradiated by CPL with the driven frequency of $\Omega=50$~meV and for different intensities in units of nm$^{-1}$. Notice that the transitions within dynamical gaps, occurring across various resonances of the system, manifest as distinct peaks or dips at lower probe frequencies. These features exhibit a marked sensitivity to occupation and relaxation mechanisms. Notably, the observed gaps prominently emerge at integer multiples of the driving frequencies. Moreover, this effect becomes more pronounced with higher intensities, aligning well with expectations derived from the underlying band structure and density of states. The legend is the same for two figures and the black dashed line in (a) represents $\mathcal{A}=0$. }
\label{Fig5}
\end{figure}

In Fig.~\ref{Fig5}(a), we illustrate the real component of the time-averaged longitudinal optical conductivity of BLG under irradiation with CPL of frequency $\Omega=50$~meV at various intensities. We focus on the regime where $\omega>\Omega$ because the optical conductivity at low energies is highly influenced by the scattering mechanisms \cite{2011Zhou}. Additionally, transitions occurring in dynamical gaps across numerous resonances of the system give rise to distinct peaks or dips at lower probe frequencies, exhibiting a strong sensitivity to occupation and relaxation mechanisms.
The gaps observed in Fig.~\ref{Fig5}(a) emerge prominently at integer multiples of the driving frequencies, and this effect becomes more pronounced with higher intensities, aligning with expectations derived from the band structure and density of states. With increasing irradiation intensity, the optical conductivity at lower frequencies decreases due to deviations from complete occupation (0 or 1) of states at low $k$ values. Moreover, some non-uniformity is noticeable in the optical conductivity at higher intensities, stemming from the fact that elevated irradiation intensities result in an increased DOS with more van Hove singularities.

In Fig.~\ref{Fig5}(b), we depict the imaginary part of the time-averaged optical Hall conductivity for various intensities of the driving field. This component is absent in the absence of a driving field. With increasing intensities of the circularly polarized driving field, the significance of this component grows, originating from the induced Berry curvature caused by the drive and indicating the presence of circular dichroism. A proof for the relationship between $Im \sigma_{xy}(\omega)$ and the Berry curvature is provided in Appendix \ref{appendix}. As we will discuss, the Berry curvature attains its maximum values for two out of the four Floquet bands, specifically in regions where the states undergo an anti-crossing. As illustrated in Fig.~\ref{Fig5}(b), the highest values of $\text{Im}\sigma_{xy}$ appear near the dynamical gaps, where the Berry curvature reaches its peak. 
The inversion of the sign in $\text{Im}\sigma_{xy}$ at low probe frequencies is linked to the shift in the physical weights of the electron-like and hole-like bands, reflecting the occupation of Floquet states at low momentum values (refer to Fig.~\ref{Fig3}(a))  

\begin{figure}
\includegraphics[width=\linewidth]{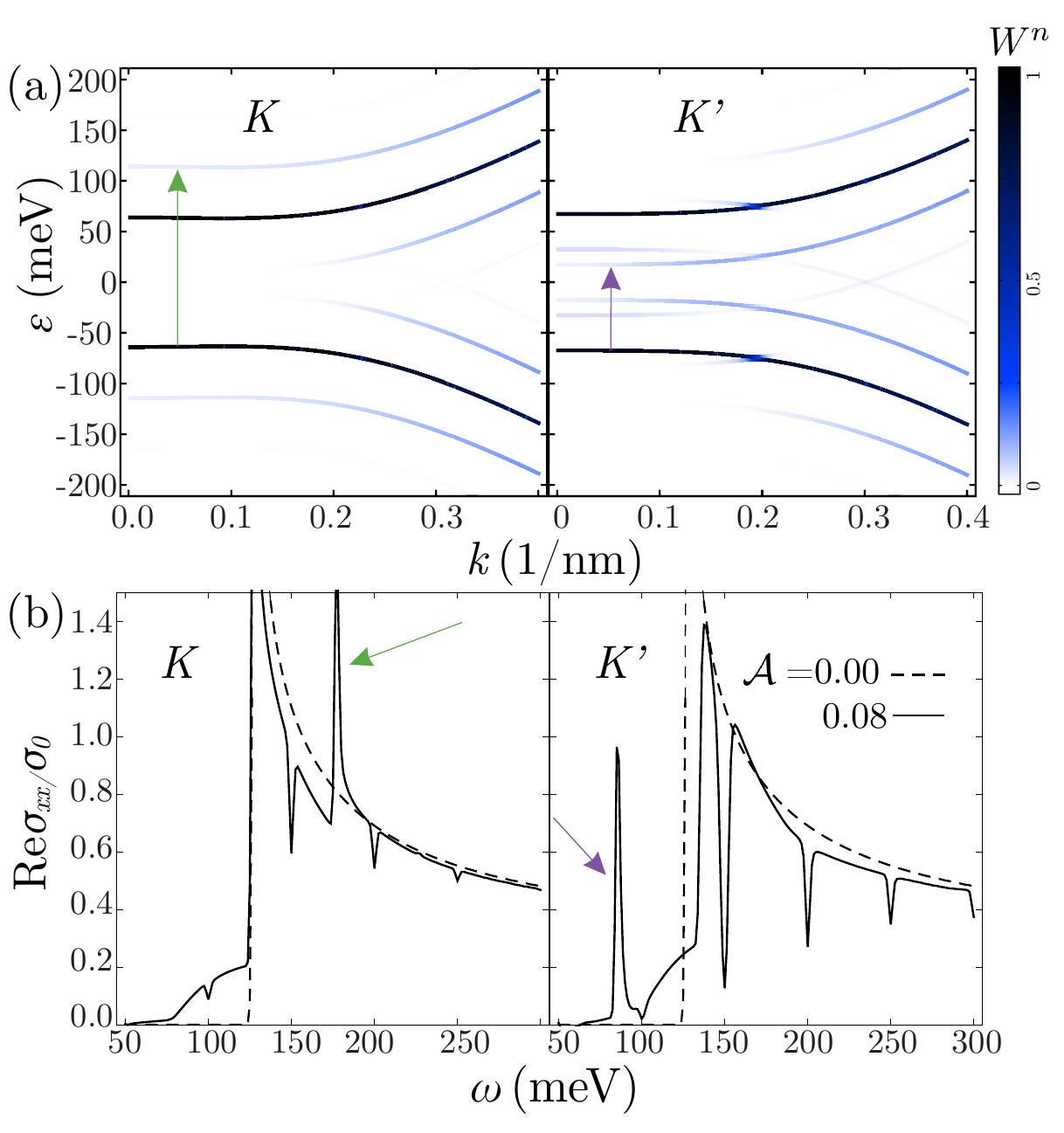}
\caption{ (Color online) (a) Quasi-energy spectrum and (b) real part of the time-averaged optical conductivity of irradiated BLG with right-handed CPL of strength
$ \mathcal{A}=0.08$~nm$^{-1}$ and frequency $ \Omega=50$~meV for applied bias $ V=135$~meV. \added{Color scale in part (a) shows the physical weights $W^n_\nu$ defined in Eq.~(\ref{weight}).} The results are presented for each valley $K, K^{'}$ separately indicating that the effective gap of the most probable Floquet bands at the K′ valley surpasses that of the K valley. The total optical conductivity is the sum of two valleys.}
\label{Fig6}
\end{figure}

We also investigate the influence of a bias potential on quasi-energies and optical conductivity. In the absence of a dynamical field ($\mathcal{A}=0$), the application of a perpendicular bias $V$ introduces a tunable band gap at $\varepsilon=0$. In contrast, in driven systems ($\mathcal{A}\neq0$), the bias potential induces an asymmetry in the dispersions of distinct valleys. Specifically, the impact of right-handed circularly polarized light (CPL) on the $K$ valley mirrors that of left-handed CPL on the $K'$ valley. Given that contributions from both valleys must be integrated in the computation of optical conductivity, right-handed and left-handed dressing lights result in identical longitudinal optical conductivity.
Fig.~\ref{Fig6}(a) presents the quasi-energy bands and time-averaged optical conductivity for various valleys. Here, a bias potential of $V=135$ meV is applied, in conjunction with right-handed polarized light of intensity $\mathcal{A}=0.08$ and frequency $\Omega=50$ meV directed towards the system. The evident asymmetry between the two valleys is conspicuous. Notably, two prominent optical transitions are denoted by arrows in the upper panel, accompanied by their respective peaks in the optical conductivity.
Furthermore, the figure indicates that the effective gap of the most probable Floquet bands at the $K'$ valley surpasses that of the $K$ valley. This leads us to infer that, at specific probe frequencies, the incident light may be primarily absorbed by one of the two valleys. Additionally, with the activation of the dressing field, multiple gaps emerge at frequencies $\omega=n\Omega$ in the optical conductivity. Noteworthy is the absorption at frequencies lower than $V$, which is an outcome of the dressing field and the introduction of new transitions between sidebands. It is imperative to highlight that the total optical conductivity is derived by aggregating the contributions from both valleys.

\subsubsection{On-resonant: Linear polarization}

\begin{figure}
\includegraphics[width=\linewidth]{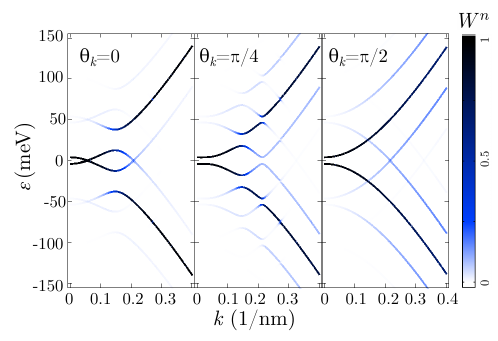}
\caption{(Color online) Quasi-energy spectrum as a function of $k $ along the direction \added{$\theta_k=\arctan(k_y/k_x)$} for the irradiated BLG by LPL with frequency $\Omega=50$~meV and intensity $\mathcal{A}=0.08$~nm$^{-1}$. \added{The color scale  shows the physical weights $W^n_\nu$ defined in Eq.~(\ref{weight}}). We observe the presence of a semimetallic structure in the most probable Floquet bands particularly along the $x$ direction. }
\label{Fig7}
\end{figure}

We now turn our attention to the influence of LPL on the system. Fig.~\ref{Fig7} presents the quasi-energy bands of BLG subjected to irradiation with LPL characterized by a frequency of $\Omega=50$~meV and an intensity of $\mathcal{A}=0.08$~nm$^{-1}$. The plot displays results for three distinct azimuthal angles in momentum space, specifically \added{$\theta_k=\arctan(k_y/k_x)=0,\pi/4,\pi/2$}, revealing a discernible anisotropy in quasienergies.

The findings in Fig.~\ref{Fig7} suggest the presence of a semimetallic structure in the most probable Floquet bands. Particularly noteworthy is the observation of a gapless point along the $x$ axis for the most probable bands, a feature not observed in other directions.

\begin{figure}
\includegraphics[width=\linewidth]{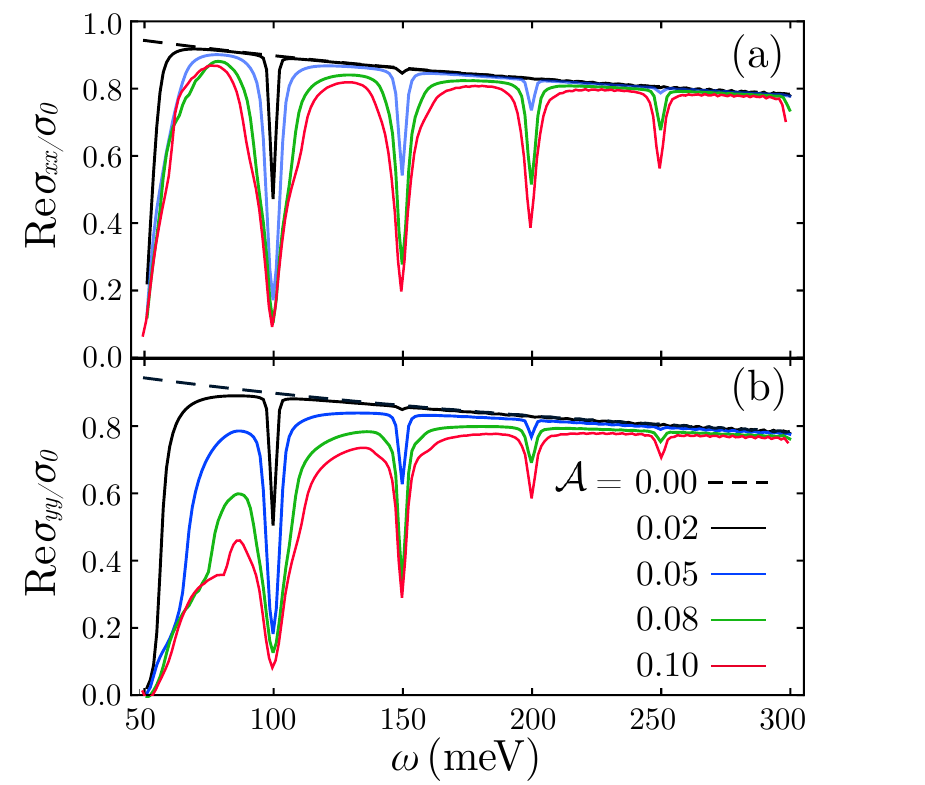}
\caption{(Color online) Real part of the time-averaged longitudinal optical conductivity of BLG irradiated by LPL with the frequency of $\Omega=50$~meV and different intensities in units of nm$^{-1}$.}
\label{Fig8}
\end{figure} 
The real component of the time-averaged longitudinal optical conductivity for irradiated BLG with LPL is depicted in Fig.~\ref{Fig8}. It is important to note that in the presence of LPL, the band dispersion becomes anisotropic, leading to distinct optical conductivities along different directions, i.e., $Re \sigma_{xx} \neq Re \sigma_{yy}$. As observed, for higher irradiation intensities and at low probe frequencies, $Re \sigma_{xx}$ surpasses. This gives rise to a linear dichroism, akin to the off-resonant case.

\section{Conclusion}\label{Sec4}

Utilizing the Floquet-Bloch theorem at zero temperature, we have studied the band spectrum, density of states, and optical conductivity of driven Bilayer Graphene (BLG) under a perpendicular applied bias. The driving field encompasses both circularly and linearly polarized lasers, acting as a dressing field for photon-assisted carriers. Our investigation has shed light on the combined impact of laser illumination with varying polarizations and applied bias potential on the band structure, state occupation, and optical conductivity.

The introduction of an applied bias disrupts the valley symmetry in the dressing spectrum induced by either right-handed circularly polarized light or its left-handed polarization. Consequently, the band structure changes with increasing applied bias for both of these polarizations. No equivalence is observed in the spectrum around the K and K′ valleys when circularly polarized light and a perpendicular bias voltage are simultaneously applied to BLG. Although the quasi-energy spectrum varies for left-handed circularly polarized and right-handed polarization, the longitudinal optical conductivity remains consistent for both polarizations. However, distinctions arise in the transversal optical conductivity. Furthermore, some sidebands emerge in the case of circularly polarized light. 

The distinct gap sizes for each valley introduce the potential for generating valley-polarized current and enabling valley-selective pumping. Additionally, circularly polarized light acting as a dressing field can induce Berry curvature, especially close to the dynamical gaps, leading to circular dichroism. Indeed, nonzero values of the imaginary part of the transverse optical conductivity which is proportional to the Berry curvature, suggest the presence of circular dichroism. Furthermore, our findings illustrate that linearly polarized light can induce anisotropy in the system, resulting in a linear dichroism in both off-resonant and on-resonant regimes.
\begin{acknowledgments}
We would like to thank F. Parhizgar for the fruitful discussions.
\end{acknowledgments}
\appendix

\section{Essentials of Floquet theorem}\label{ftheorem}
 The time-dependent Schr\"{o}dinger equation (working in lattice $k$-space and omitting the $k$ index) is given by: $$i \partial_{t} \vert \psi_{\nu}(t)\rangle=\hat{H}(t) \vert \psi_{\nu}(t)\rangle,$$ where $\nu$ denotes the band index. For a time-periodic Hamiltonian of the form:
$$ \hat{H}(t+T)=\hat{H}(t),~T=2\pi/\Omega,$$ it can be solved using the Floquet approaches \cite{1973Sambe}.

Let us write the solution of the Schr\"{o}dinger equation based on the Floquet states,
$ |\psi_\nu(t)\rangle=e^{-i\varepsilon_\nu t}|\phi_\nu(t)\rangle $ where $ |\phi_\nu(t+T)\rangle=|\phi_\nu(t)\rangle $ is time-periodic Floquet quasi mode, and the Schr\"{o}dinger solution obeys a Bloch-type theorem 
$$|\psi_\nu(t+T)\rangle=e^{-i\varepsilon_\nu T }|\psi_\nu(t)\rangle .$$ The Floquet quasi mode
$ |\phi_\nu(t)\rangle$ is an eigenstate of operators $ (\hat{H}(t)-i \partial_{t})$
\begin{equation}
\left[ \hat{H}(t)-i\partial_{t} \right] |\phi_\nu(t)\rangle =\varepsilon_{\nu} |\phi_\nu(t)\rangle. 
\label{hf1}
\end{equation}
where $\varepsilon_{\nu}$ are quasi energies.
Since $ \phi_{\nu}(t) $ is time-periodic, it can be expanded by using the Fourier series,
\begin{equation}
|\phi_\nu(t)\rangle=\sum_{n}e^{-in\Omega t}|\phi^{(n)}_\nu \rangle
\label{phifourier}
\end{equation}
where $ \nu=1,2,3,4 $ represents the band index, and $|\phi^{(n)}_\nu \rangle$ denote the time-independent expansion coefficients, which form a complete basis set in the Hilbert space of $ \Re \bigotimes T$. The orthonormality condition \cite{1965Shirley, 1973Sambe} is given by $ \langle\langle \phi_{\nu}\vert\phi_{\beta} \rangle\rangle=\dfrac{1}{T}\int_{0}^{T} \langle\phi_{\nu}(t)\vert\phi_{\beta}(t) \rangle dt=\delta_{\nu\beta}$.
\\
By performing a Fourier expansion on $|\phi_\nu(t)\rangle$ as described in Eq.~(\ref{phifourier}), and similarly expanding the Hamiltonian as 
\begin{equation}
\hat{H}(t)=\sum_{n}e^{-in\Omega t}\hat{H}^{(n)},
\label{ht}
\end{equation}
 we can express Eq.~(\ref{hf1}) in the following matrix form \cite{handbook}: 
\begin{equation}
(\varepsilon_\nu+m  \Omega )|\phi_\nu^{(m)}\rangle=\sum_{m'}H^{(m-m')}|\phi_\nu^{(m')}\rangle
\label{hf}
\end{equation}
Given that the quasienergies $\varepsilon_\nu$ are defined modulo $n\Omega$, we introduce the reduced quasi energy $\epsilon_\nu$ to lie within the first Floquet-Bloch zone $(-\Omega/2,\Omega/2)$.
The size of the Floquet Hamiltonian matrix is $ n_F\times n_F $, where $ n_F=(2n_t+1)n_{cell} $. Here, $2 n_t+1 $ represents the number of replicas, and $ n_{cell}=4 $ stands for the number of unirradiated BLG energy bands. The corresponding weights of subbands in this context can be derived from \cite{2011Zhou}.
\begin{equation}
W_{\nu}^n= \langle \phi^{(n)}_{\nu}|\phi^{(n)}_{\nu}\rangle
\label{weight}
\end{equation} 
Another important physical parameter we define is the mean energy which is single-valued for all $ \varepsilon_{\nu}^{n} $ and is defined as
\cite{Dittrich, 2013Scholz,1997Faisal} 
\begin{equation}
\begin{aligned}
\bar{\epsilon}_{\nu}&=\dfrac{1}{T}\int_{0}^{T} dt \left\langle \psi_{\nu}(t)|\hat{H}(t)|\psi_{\nu}(t)\right\rangle \\
&=\epsilon_{\nu}+\sum_{n}n  \Omega W_{\nu}^n
\label{menergy}
\end{aligned}
\end{equation}
where $ n $ refers to Floquet subbands. Also, the time-averaged density of states in the Floquet picture is derived as \cite{2009Oka,2011Zhou, 2013Scholz}
\begin{equation}\label{DOSF}
{\text {DOS}(E)}= \sum_{\mathbf{k} \nu n} g_s W_{\nu}^n
\delta(E-(\epsilon_{\nu }+n  \Omega))
\end{equation}
where $ g_s=2$ shows the spin degeneracy.

\section{Imaginary part of optical Hall conductivity and Berry curvature}\label{appendix}
As explained in the main text, the imaginary part of the optical Hall conductivity has a close relationship with Berry curvature. In this Appendix, we explicitly prove it by rewriting Eq.~(\ref{ochall}) in the following form \cite{fiete2016}
\begin{equation}
\begin{aligned}
&\text{Im} {\bar{\sigma}_{ll'}(\omega)}=\pi \omega g_s\sum_{\textbf{k},m}\sum_{\nu<\mu} 
 F_{\nu\mu}^m  (f_{\nu}-f_{\mu})\times\\
&\big[\delta(\omega+\epsilon_{\nu}-\epsilon_{\mu}-m\Omega)+\delta(\omega-(\epsilon_{\nu}-\epsilon_{\mu}-m\Omega))\big]
\end{aligned}
\label{a1eq}
\end{equation}
where 
\begin{equation}
\begin{aligned} 
 F_{\nu\mu}^m=\frac{\text{Im}[ j^{l(m)}_{\nu\mu} j^{l'(-m)}_{\mu\nu}]}{(\epsilon_\nu-\epsilon_\mu-m\Omega)^2}.
\end{aligned}
\label{}
\end{equation}
We call $F_{\nu\mu}^m$ the $m$-photon component of the Berry curvature as the mixing states of the band $\nu$ with the band $\mu$. 
The total Berry curvature of a band can be written as: 
\begin{equation}
\begin{aligned} 
 \boldsymbol{\Omega}_\nu=\sum_m\sum_{\mu\neq \nu}F_{\nu\mu}^m.
\end{aligned}
\label{a3eq}
\end{equation}

Subsequently, it is implied by Eq.~(\ref{a1eq}) that the components of the Berry curvature, weighted by the difference between the occupation of Floquet states, determine the imaginary part of the optical Hall conductivity. 

\begin{figure}
\includegraphics[width=\linewidth]{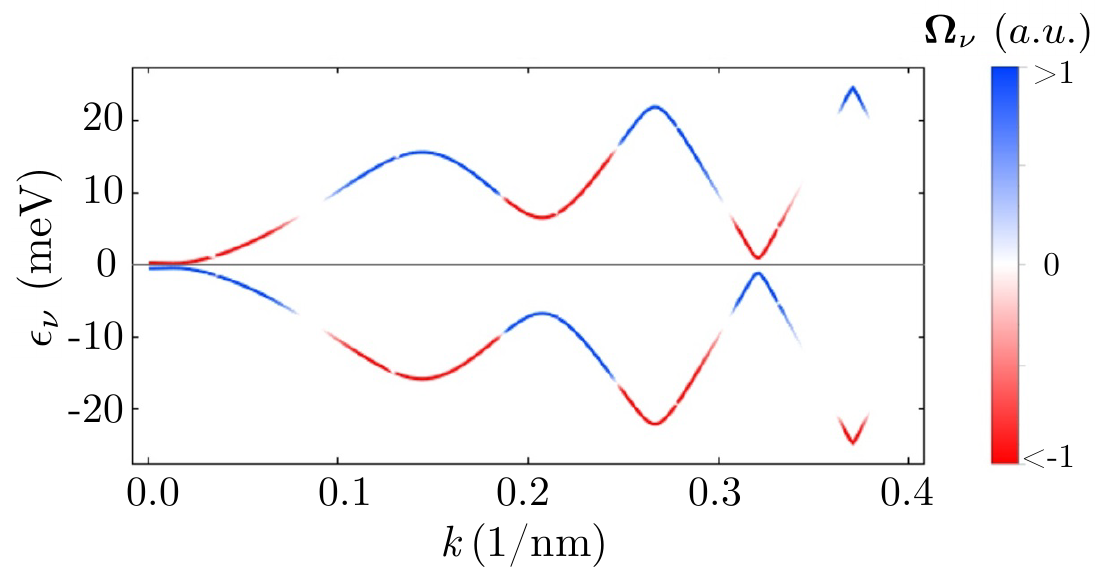}
\caption{(Color online) Two bands of four quasi-energy bands in the first Floquet-Brillouin zone for irradiated BLG with CPL with frequency of $\Omega=50$~meV and intensity $\mathcal{A}=0.08$~nm$^{-1}$. The color on each band indicates the Berry curvature calculated from Eq.~(\ref{a3eq}).}
\label{Fig9}
\end{figure} 
We present the Berry curvature as a color-scaled bar overlaid on the quasi-energy bands of BLG irradiated by CPL in Fig.\ref{Fig9} (this can be compared with Fig.\ref{Fig4}(a) where each Floquet band is depicted in distinct colors). It is observed that only two out of four Floquet bands exhibit a non-negligible amount of Berry curvature, with this curvature reaching its maximum values at light-induced anti-crossing states that emerge following a mixing between two Floquet bands. In the absence of light, there is no Berry curvature in the bands, and consequently, no circular dichroism.

\end{document}